# Effective reaction temperatures of irreversible dust chemical reactions in a protoplanetary disk

Lily Ishizaki[1,2], Shogo Tachibana[2], Tamami Okamoto[3], Daiki Yamamoto[4], Shigeru Ida[3]

Running head: Effective reaction temperatures of dust chemical reactions

## ABSTRACT

Dust particles in protoplanetary disks experience various chemical reactions under different physicochemical conditions through their accretion and diffusion, which results in the radial chemical gradient of dust. We performed three-dimensional Monte Carlo simulations to evaluate the dust trajectories and the progress of fictitious irreversible reactions, of which kinetics is expressed by the Johnson-Mehl-Avrami equation. The distribution of the highest temperature that each particle experiences before the degree of reaction exceeds a certain level shows the log-normal distribution, and its mode temperature was used as the effective reaction temperature. Semi-analytical prediction formulas of the effective reaction temperature and its dispersion were derived by comparing a reaction timescale with a diffusive transport timescale of dust as a function of the reaction parameters and the disk parameters. The formulas reproduce the numerical results of the effective reaction temperatures and their dispersions within 5.5 and 24 %, respectively, in a wide temperature range (200–1400 K). We applied the formulas for the crystallization of amorphous silicate dust and its oxygen isotope exchange with the $H_2O$ vapor based on the experimentally determined kinetics. For sub-micron sized amorphous forsterite dust, the predicted effective reaction temperature for the oxygen isotope exchange was lower than that of crystallization without overlap even considering their dispersions. This suggests that the amorphous silicate dust in the protosolar disk could exchange their oxygen isotopes efficiently with the $^{16}$O-poor $H_2O$ vapor, resulting in the distinct oxygen isotope compositions from the Sun.

---

[1] Corresponding author r.ishizaki@eps.s.u-tokyo.ac.jp
[2] Department of Earth and Planetary Science, The University of Tokyo, Hongo, Tokyo 113-0033, Japan
[3] Earth-Life Science Institute, Tokyo Institute of Technology, Meguro, Tokyo 152-8550, Japan
[4] Department of Earth and Planetary Sciences, Kyushu University, Motooka, Nishi-ku, Fukuoka 819-0395, Japan

# 1. INTRODUCTION

Various dust chemical reactions occur in a protoplanetary disk depending on physicochemical conditions that dust experiences through its accretional and diffusive transport. Such chemical reactions would be responsible for making the chemical diversity of dust in the disk.

Chemical reactions of dust moving in protoplanetary disks have been modeled using Monte Carlo simulations to simulate the diffusive random motion of dust (e.g., Ciesla 2011; Ciesla & Sandford 2012; Okamoto & Ida 2022). Ciesla (2011) developed the Monte Carlo simulation of dust particles movement in an accreting protoplanetary disk to show that crystalline silicate particles initially at ~ 5 au can diffuse out to the region beyond 20 au, suggesting that dust experiencing a certain temperature for complete crystallization can be incorporated into comets that formed well beyond the snow line. With the same 3D Monte Carlo simulation, Ciesla and Sandford (2012) demonstrated that amorphous ice dust is stirred up above the disk photo-surface long enough and accumulation of radicals in the dust by UV photon irradiation produces complex organic molecules in a protoplanetary disk.

Okamoto and Ida (2022) discussed the redistribution of silicate dust particles that crystallize at the crystallization line (~1000 K) in a protoplanetary disk using the 3D Monte Carlo simulation. Considering the pebble accretion model (e.g., Lambrechts & Johansen 2014) and assuming that the pebbles consist of many small amorphous silicate particles embedded in the icy mantle, they adopted the simulation setting that the small-sized silicate particles are released by ice sublimation at the snow line (e.g., Saito & Sirono 2011; Morbidelli et al. 2015; Ida & Guillot 2016). Because they consistently simulated the trajectories of both particles that once moved into the region with > 1000 K to be crystallized and those staying in amorphous structure, they quantitatively calculated crystallization ratio of silicate dust particles that diffuse out to the outer region of the disk to be incorporated in comets, although the estimated fraction of crystalline dust is a few times smaller than the observation, as long as the amorphous silicate dust is continuously supplied by icy pebbles.

These previous studies focused on specific reactions of dust in the accretion disk, but various dust chemical reactions occur under protoplanetary disk conditions, which should contribute to the chemical evolution of the disk. The difference in critical reaction temperatures should thus be responsible for making the radial chemical gradient of dust. Because the dust moves advectively and diffusively in the accretion disk, the chemical reaction of dust should be discussed with its kinetics that has a strong temperature dependence.

In this work, we aim to determine critical reaction temperatures of various irreversible chemical reactions of dust in a steady accretion disk and obtain the general expression of the reaction temperatures as a function of both reaction and disk parameters.

In the next section, we describe the protoplanetary disk model, the 3D Monte Carlo simulation method to track trajectories of dust particles and their chemical reactions. Section 3 shows the results of the simulations focusing on the highest temperature that dust experience until the reaction degree reaches a certain level. Section 4 discusses the semi-analytical general formula of the effective reaction temperature. We apply the formula for the reaction temperatures of crystallization and oxygen isotope exchange of amorphous silicate dust and discuss its cosmochemical implication in Section 5, followed by the conclusion in Section 6.

## 2. METHOD

### 2.1. Disk Model

In this work, we simulated irreversible chemical reactions of dust particles moving advectively and diffusively in a steady accretion disk around a solar-mass central star ($M_\odot$) using the 3D Monte Carlo simulation (e.g., Ciesla 2011; Ciesla & Sandford 2012; Okamoto & Ida 2022).

We adopted a steady accretion disk with a mass accreting rate of $\dot{M}$. Diffusivity of dust, $D$, is given as the gas turbulent viscosity, $\nu_{\rm vis} = \alpha c_s H$, where $\alpha$ is a dimensionless viscous parameter ($0 < \alpha \leq 1$), $H$ is the local gas scale height, and $c_s$ is the local speed of sound (the $\alpha$-viscosity model; Shakura & Sunyaev 1973).

The accreting disk is heated by uniform viscous heating, proportional to the local gas spatial density. We assumed that the disk is in hydrostatic equilibrium such that $\rho(z) = (\Sigma_g/\sqrt{2\pi} H) \exp(-z^2/2H^2)$ in the vertical direction, where $\Sigma_g$ is the disk gas surface density and $z$ is the height. The optical depth at $z$ is calculated as

$$\tau_z \equiv \int_{|z|}^{\infty} \kappa \rho(z) dz = \frac{\kappa \Sigma_g}{\sqrt{\pi}} \int_{z/\sqrt{2}H}^{\infty} \exp(-x^2)\, dx = \frac{\kappa \Sigma_g}{2}\left[1 - {\rm erf}\left(\frac{z}{\sqrt{2}H}\right)\right], \quad (1)$$

where $\kappa$ is the opacity, and ${\rm erf}(x)$ is the error function:

$$ {\rm erf}(x) \equiv \frac{2}{\sqrt{\pi}} \int_x^{\infty} \exp(-x'^2)\, dx'. \quad (2)$$

The opacity is assumed to be $\kappa = 2.5$ cm$^2$/g (e.g., Pollack et al., 1985, 1994). We here note that the value of $\kappa$ changes the thermal structure of the disk, but we found that it does not affect the discussion on the effective reaction temperatures in the range of $\kappa$=1–4 cm$^2$/g.

The photo-surface temperature of the disk is given by

$$T_{\rm surf} \simeq 85 \left(\frac{\dot{M}}{10^{-8}\,M_\odot/{\rm yr}}\right)^{1/4} \left(\frac{r}{1{\rm au}}\right)^{-3/4}. \quad (3)$$

and temperature at z is given by

$$T(z) = \left[1 + \frac{3(\tau_z - 1)}{4}\left(1 - \frac{\tau_z - 1}{\tau_{\rm disk}}\right)\right]^{1/4} T_{\rm surf}, \quad (4)$$

where $\tau_{\rm disk}$ is the disk optical depth, defined by

$$\tau_{\rm disk} \equiv \int_{-\infty}^{\infty} \kappa \rho(z)\, dz. \quad (5)$$

The derivation of Eq. (4) is given in Appendix. In the analytical derivation of the midplane temperature from $T_{\rm surf}$, it is often assumed that the viscous heating is concentrated at $z = 0$ (Lin & Papaloizou 1985). However, because chemical reactions generally have sensitive temperature dependences, we use the disk temperature as a continuous function of $z$ given by Eq. (4) with the assumption that viscous heating rate is proportional to $\rho(z)$. The temperature gradient along the $z$ direction was not included in previous studies (Ciesla 2011; Ciesla & Sandford 2012; Okamoto & Ida 2022), and its effect on the reaction progress will be discussed below.

The steady mass accretion rate is given by $\dot{M} \simeq 3\pi \Sigma_g \nu_{\rm vis}$, and the gas surface density is expressed as

$$\Sigma_g \simeq 165 \left(\frac{\alpha}{10^{-2}}\right)^{-4/5} \left(\frac{\dot{M}}{10^{-8}M_\odot \text{ yr}^{-1}}\right)^{3/5} \left(\frac{r}{1\text{au}}\right)^{-3/5} \text{ g/cm}^2. \tag{6}$$

At the midplane where $z = 0$ and $t_z = t_{disk}/2 = \kappa\Sigma_g/2$, the disk temperature $t_{disk} \gg 1$ is given by

$$T_c \simeq \left(\frac{3\,\tau_{disk}}{16}\right)^{1/4} T_{surf} \simeq 607 \left(\frac{\alpha}{10^{-2}}\right)^{-1/5} \left(\frac{\dot{M}}{10^{-8}M_\odot \text{ yr}^{-1}}\right)^{2/5} \left(\frac{r}{1\text{au}}\right)^{-9/10}. \tag{7}$$

### 2.2. Trajectories of Dust Particles

Following Ciesla (2010, 2011), we performed 3D Monte Carlo simulations to evaluate the trajectories of dust particles in the steady-state accretion disk. Dust size was assumed to be small enough to be well coupled with gas. The 3D Monte Carlo Lagrangian simulation model reproduces the surface density evolution of dust particles following the Eulerian advection-diffusion equation (Ciesla 2010, 2011):

$$\frac{\partial \Sigma_d}{\partial t} = \frac{1}{r}\frac{\partial}{\partial r}\left[r\Sigma_d\left(v_r - \frac{D}{\Sigma_d/\Sigma_g}\frac{\partial}{\partial r}\left(\frac{\Sigma_d}{\Sigma_g}\right)\right)\right], \tag{8}$$

where $\Sigma_d$ is the surface density of dust and $v_r$ is their radial advection velocity (e.g., Okamoto & Ida 2022). The motions of an individual particle are described by

$$x_i = x_{i-1} + v_x \delta t_{phys} + \mathcal{R}_x \sqrt{6D(x')\delta t_{phys}}, \tag{9}$$

$$y_i = y_{i-1} + v_y \delta t_{phys} + \mathcal{R}_y \sqrt{6D(y')\delta t_{phys}}, \tag{10}$$

$$z_i = z_{i-1} + v_z \delta t_{phys} + \mathcal{R}_z \sqrt{6D(z')\delta t_{phys}}, \tag{11}$$

Where $\delta t_{phys}$ is the timestep to calculate trajectories, $\delta t_{phys} = \Omega_K^{-1}$, where $\Omega_K$ is the local Keplerian frequency, and $\mathcal{R}_x$, $\mathcal{R}_y$, and $\mathcal{R}_z$ are independent random numbers with a value between –1 and 1. The second and third terms on the right represent advection and diffusion, respectively. Accounting for gradients in the diffusivity, $x'$, $y'$, and $z'$ are given by $x' = x_{i-1} + (\partial D/\partial x)/2$, $y' = y_{i-1} + (\partial D/\partial y)/2$, and $z' = z_{i-1} + (\partial D/\partial z)/2$ (e.g., Visser, 1997; Ciesla, 2010). The advection velocities of dust accreting with gas in the steady accretion disk are given by

$$v_x = v_{drag,r} \times \frac{x}{r}, \tag{12}$$

$$v_y = v_{drag,r} \times \frac{y}{r}, \tag{13}$$

$$v_z = v_{drag,z} + \partial D(z)/\partial z - \alpha\Omega_K z, \tag{14}$$

where $x$, $y$, and $z$ are the values of the Cartesian coordinates, and $r$ is the radial distance from the central star, given by $r = \sqrt{x^2 + y^2}$. In this well-coupled case, the radial velocity is given by $v_{drag,r} = v_r = -3n_{vis}/2r = -(3/2)\alpha(H/r)^2 v_K$, where $v_K$ is the local Keplerian velocity. Because $v_{drag,z} = 0$ and $\partial D(z)/\partial z = \partial v_{vis}/\partial z = 0$, $v_z = -\alpha\Omega_K z$. To check the validity of the simulation, we made the same dust trajectory calculation with those in previous studies (Ciesla 2010, 2011; Okamoto & Ida 2022), and confirmed that our trajectory calculations give the consistent results with previous studies.

In the simulations, $10^4$ dust particles were released at the disk midplane near the H$_2$O snowline ($T \approx 180$ K) for each disk parameter set (see Section 2.4), according to the simulation setting adopted by Okamoto and Ida (2022). Because we do not consider collective effects, back-reaction from dust to gas nor dust growth, the dust particle trajectories are independently integrated. The calculations continued for up to $10^6$ years. We note here that this initial location has almost no effect on the estimate of the reaction temperatures as long as they occur at >200 K.

### 2.3. Reaction Kinetics

Progresses of many irreversible chemical reactions are well described empirically by the Johnson-Mehl-Avrami (JMA) equation (Avrami 1939; Johnson & Mehl 1939):

$$X = 1 - \exp\left[-\left(\frac{t}{\tau}\right)^n\right], \tag{15}$$

where $X$ is the degree of each reaction, $t$ is time, and $n$ is the Avrami index. The time evolution of $X$ is represented by the sigmoidal-shaped curve, resulting in slowing the reaction rate before completion. Therefore, this empirical equation implicitly includes the reduction of the driving force of reaction when the reaction approaches the equilibrium.

The characteristic reaction timescale, $\tau$, is given by

$$\tau = \left[\nu_{\text{chem}} \exp\left(-\frac{E_a}{RT}\right)\right]^{-1}, \tag{16}$$

where $\nu_{\text{chem}}$ is the pre-exponential factor, $E_a$ is the activation energy, and $R$ is the gas constant. The reaction parameters $n$, $\nu_{\text{chem}}$, and $E_a$ can be determined by laboratory experiments.

We note here that the dependence of kinetic parameters on the disk gas density and/or gas chemistry is not considered for simplicity although their dependence on the gas chemistry has been reported in some cases (e.g., crystallization of amorphous silicate with the composition of Mg$_2$SiO$_4$; Yamamoto and Tachibana, 2018). We do not consider reversible chemical reactions such as evaporation and condensation of dust, whose kinetics is not expressed simply by Eq. 16 (e.g., Takigawa et al., 2009; Tachibana et al., 2011; Takigawa et al., 2015), either. The driving force of the reaction (Gibbs energy change of the reaction DG) depends on the degree of progress of reversible reactions, and in some cases is a function of gas density (e.g., evaporation of minerals in hydrogen gas: Tachibana and Tsuchiyama, 1998; Tsuchiyama et al., 1999; Tachibana et al., 2002; Takigawa et al., 2009). These reactions are also important targets for future work.

We simulated the progress of fictitious chemical reactions of dust moving in the disk as described in Section 2.2. In the simulation, $X$ is set to zero when the dust is released at the snowline. For each timestep of the Monte Carlo simulation, the increment $\delta X$ is calculated by the following differentiated JMA equation with local $T$ at the position of the particle:

$$\delta X = (1 - X)\left(1 - \exp\left(-n\frac{\delta t_{\text{chem}}}{\tau}(-\ln(1-X))^{1-\frac{1}{n}}\right)\right). \tag{17}$$

Because the chemical reaction has a strong temperature dependence, the reaction may proceed significantly during the physical timestep $\delta t_{\text{phys}}$ depending on the reaction parameters and the temperature. To prevent such an abrupt reaction progress during $\delta t_{\text{phys}}$, we newly introduced the chemical timestep $\delta t_{\text{chem}}$. The chemical time step is defined as the duration for a reaction to progress with $\delta X_{\text{lim}}$ of 0.01:

$$\delta t_{\text{chem}} = -\frac{n}{\tau(-\ln(1-X))^{1-\frac{1}{n}}} \ln\left(1 - \frac{\delta X_{\text{lim}}}{1-X}\right), \tag{18}$$

and was used when $\delta X$ is larger than $\delta X_{\text{lim}}$ for a given $\delta t_{\text{phys}}$ (Section 2.2). We assumed that the dust keeps linear motion during $\delta t_{\text{phys}}$, and the local temperature was updated at each chemical time step $\delta t_{\text{chem}}$ until the sum of $\delta t_{\text{chem}}$ becomes equal to $\delta t_{\text{phys}}$. Note that $\delta t_{\text{chem}}$ is set equal to $\delta t_{\text{phys}}$ when $\delta X < \delta X_{\text{lim}}$ during $\delta t_{\text{phys}}$.

### 2.4. Parameters

The following disk parameters were used for the simulation; the viscous parameter $\alpha$ of $10^{-2}$ and $10^{-3}$ (e.g., Sano et al. 2004) and the steady accretion rate $\dot{M}$ of $10^{-6}$, $10^{-7}$ and $10^{-8}$ $M_\odot$ yr$^{-1}$ (Muzerolle et al. 2000).

In order to simulate a wide variety of chemical reactions (Eq. 16), we surveyed hypothetical reactions with the logarithm of pre-exponential factor of $\ln(\nu_{\text{chem}}[\text{s}^{-1}])$ of 10, 20, 30, 40, 50, and 60, the Avrami index $n$ of 0.5, 1.0, 1.5, 2.5, and 4.0, and the activation energy $E_a$ of 20, 50, 100, 150, 200, 300, 400, 500, 600, 800, and 1000 kJ mol$^{-1}$. These parameters cover various reactions occurring in a wide temperature range (e.g., desorption of chemisorbed molecules, volume diffusion of elements in solids, crystallization of amorphous silicates, gas-solid reactions, and evaporation of minerals) (Tachibana & Tsuchiyama, 1998; Tachibana et al., 2002; Takigawa et al., 2009; Tachibana et al., 2011; Takigawa et al., 2015; Yamamoto & Tachibana, 2018; Kuroda et al., 2018, 2019; Yamamoto et al., 2018, 2020, 2021). We assumed that the reactions are not dependent on the disk gas pressure in this study.

The effective reaction temperature of a chemical reaction is discussed below using the highest temperature ($T_{\text{max}}^{X_{\text{reac}}}$) that each dust particle experiences before the degree of reaction reaches a certain level $X_{\text{reac}}$. The $T_{\text{max}}^{X_{\text{reac}}}$ at $X_{\text{reac}}$ of 0.8, 0.9 and 0.99 were recorded for each dust in this study. Because chemical reactions are the most sensitive to temperature as shown in Eq. (16), we are concerned with the highest temperature.

By combining these parameters, we made 5940 3D Monte Carlo simulations (330 reactions × 6 disk parameters × 3 $X_{\text{reac}}$).

# 3. NUMERICAL RESULTS
## 3.1. Progress of a chemical reaction of dust in the accreting disk

A representative trajectory of a dust particle is shown in Fig. 1(a). The overall trend of dust movement is the inward advection due to accretion, but the diffusive movement due to turbulence is overprinted. In some cases, the outward diffusion of dust also occurs.

Progress of a chemical reaction ($\ln(\nu_{chem}[\text{s}^{-1}]) = 30$, $E_a = 300$ kJ mol$^{-1}$ and $n = 0.5$) of the same particle is shown in Fig. 1(b) with the temperature that the dust experiences until $X$ reaches 0.99. It is found that the reaction begins to proceed when the dust moves to the region where the temperature is ~600 K. Further advective motion of the dust brings it into the inner higher temperature region, leading to the progress of the reaction. Some horizontal lines in Fig. 1(b) represent a halt of the reaction due to diffusion of the dust upward/downward or outward into the lower temperature region, where the strong temperature dependence of the reaction rate prevents the progress of the reaction. The reaction effectively proceeds when the dust is in the region with temperature of 600-800 K (Fig. 1(b)). The highest temperature that the dust experiences before $X$ reaches $X_{\text{reac}} = 0.99$ ($T_{\max}^{0.99}$) is 793 K, shown with a vertical line in Fig. 1(b).

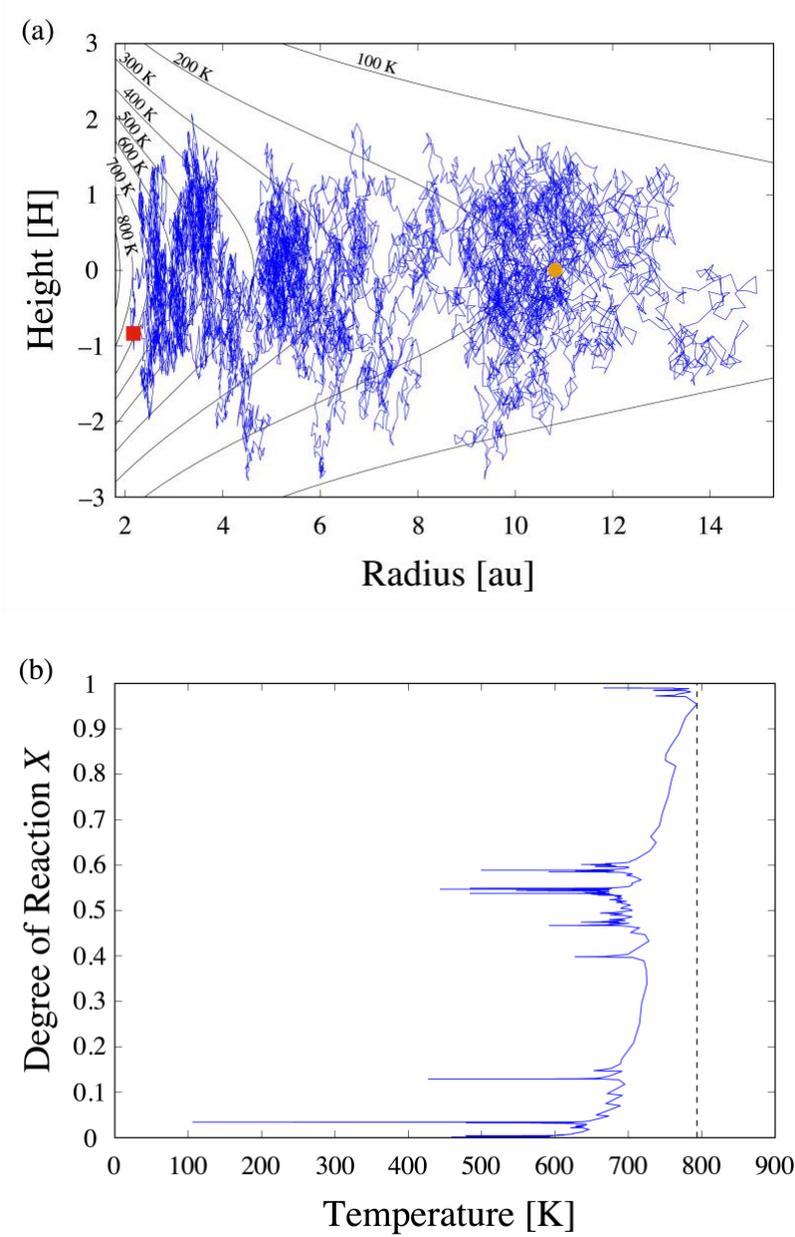

**Figure 1.** (a) A representative path of a particle that moves in a steady accretion disk with $\alpha=10^{-2}$ and $\dot{M} = 10^{-6}\ M_\odot\ \mathrm{yr}^{-1}$. The vertical axis is the disk height $z$ normalized to the scale height $H$ at a given radius. Contour lines show the disk temperature. The particle was released at the orange circle point and completed a reaction ($X_{\mathrm{reac}} = 0.99$) at the red square point. (b) A progress of the reaction and a temperature profile of the dust in Fig. 1(a). The reaction parameters are $\ln(\nu_{\mathrm{chem}}[\mathrm{s}^{-1}]) = 30$, $E_a = 300\ \mathrm{kJ\ mol^{-1}}$ and $n = 0.5$. A vertical line shows the highest temperature $T_{\mathrm{max}}^{0.99}$ that the dust experiences before $X$ reaches $X_{\mathrm{reac}}$ of 0.99 (793 K).

### 3.2. Effective reaction temperatures and their dispersion

We made histograms of $T_{\max}^{X_{\text{reac}}}$ for the particles, of which $X$ exceeds $X_{\text{reac}}$ during the simulation. The bin width of the histogram was determined by Scott's choice as $3.5\hat{\sigma}/\sqrt[3]{N}$, where $\hat{\sigma}$ is the standard deviation of $T_{\max}^{X_{\text{reac}}}$ and $N$ is the total number of particles (Scott, 1979). Some examples of the histograms are shown in Fig. 2.

All the $T_{\max}^{X_{\text{reac}}}$ histograms have asymmetric tails extending toward the higher temperature (Fig. 2), and they are well fitted by the log-normal distribution:

$$\frac{1}{\sqrt{2\pi}\sigma T} \exp\left\{-\frac{(\log T - \mu)^2}{2\sigma^2}\right\}. \tag{22}$$

The mode temperature of the $T_{\max}^{X_{\text{reac}}}$ distribution is defined here as the effective reaction temperature ($T_{\text{reac}}$) for the reaction and is expressed as

$$T_{\text{reac}} = \exp(\mu - \sigma^2). \tag{23}$$

The relative dispersion of $T_{\text{reac}}$ is given as $\sigma_{\text{reac}} = \left\{e^{2\mu+\sigma^2}(e^{\sigma^2}-1)\right\}^{1/2}/T_{\text{reac}}$. When $\sigma_{\text{reac}} \ll 1$, $\sigma_{\text{reac}}$ nearly equals $\sigma$ in Eq. (22). We found that $\sigma_{\text{reac}}$ in all the simulations can be represented by $\sigma$. For instance, in the case of Fig. 2(a), $T_{\text{reac}}$ of 1312 K and $\sigma_{\text{reac}}$ of 0.0275 (<<1) are obtained.

The effective reaction temperatures ($T_{\text{reac}}$) and their relative dispersion ($\sigma_{\text{reac}}$) for various fictitious reactions vary depending on the reaction parameters and disk parameters (Fig. 3). We here note that the simulation results with $T_{\text{reac}}$ of >1500 K are not used for further discussion because silicate dust sublimates at such high temperatures and the assumption of $\kappa$=2.5 cm$^2$/g (Section 2) is no longer valid. The results with $T_{\text{reac}}$ of < 180 K or those with the $T_{\max}^{X_{\text{reac}}}$ distribution with a tail down to <180 K are also excluded because the dust release temperature was ~180 K in the numerical simulations, and the effect of the dust release location cannot be ignored. With those exceptions, $T_{\text{reac}}$ obtained in the present simulation ranges from 188 to 1499 K, which covers various chemical reactions occurring in a wide range of temperatures.

As a natural consequence of temperature-dependent reactions, $T_{\text{reac}}$ becomes larger as $E_a$ increases or $\ln(\nu_{\text{chem}}[\text{s}^{-1}])$ decreases because the reactions require higher temperatures to proceed effectively with larger $E_a$ and smaller $\nu_{\text{chem}}$ (Fig. 3). The $\sigma_{\text{reac}}$ increases as $\ln(\nu_{\text{chem}}[\text{s}^{-1}])$ decreases but it does not depend largely on $E_a$ (Fig. 3). The $T_{\text{reac}}$ and $\sigma_{\text{reac}}$ show weak dependence on the Avrami index $n$. They increase as $n$ decreases (Fig. 3) because a reaction occurs effectively when $X$ exceeds $1 - (1/e) \approx 0.632$ for a larger $n$.

As for the dependence on disk parameters, $T_{\text{reac}}$ shows positive and negative correlation with $\alpha$ and $\dot{M}$, respectively, while $\sigma_{\text{reac}}$ has little dependence on the disk parameters. The origins of these dependences/independences are shown through the derivation of the generalized formula in the next section.

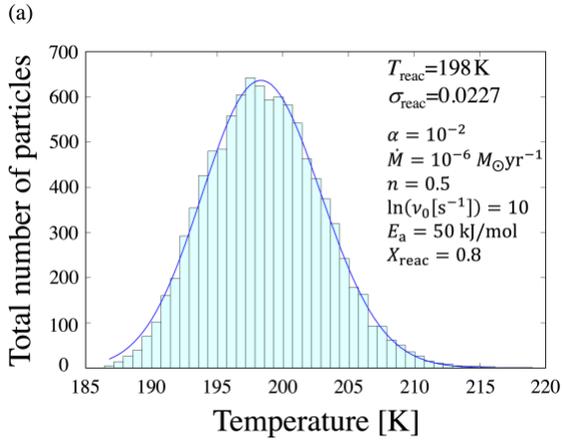
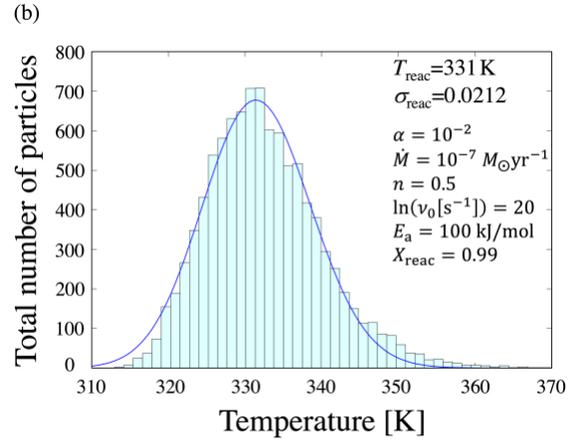
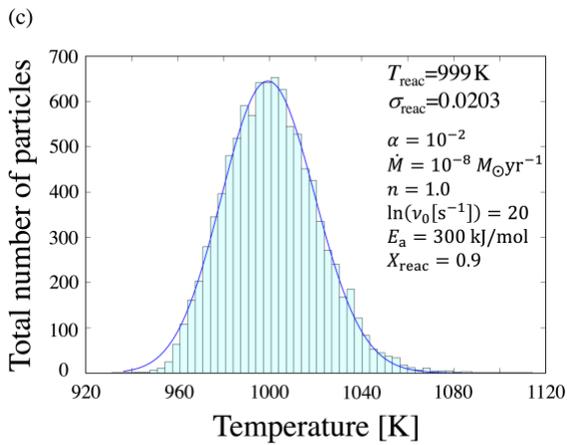
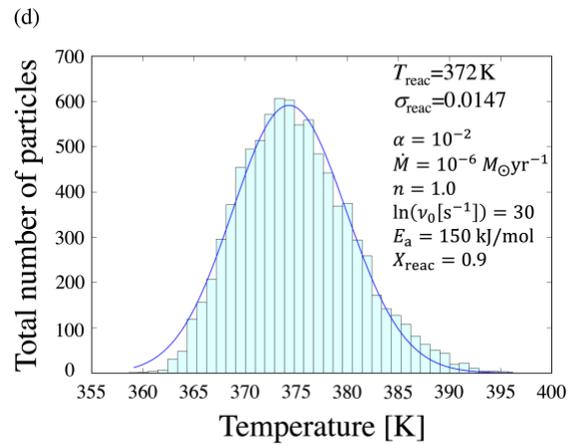
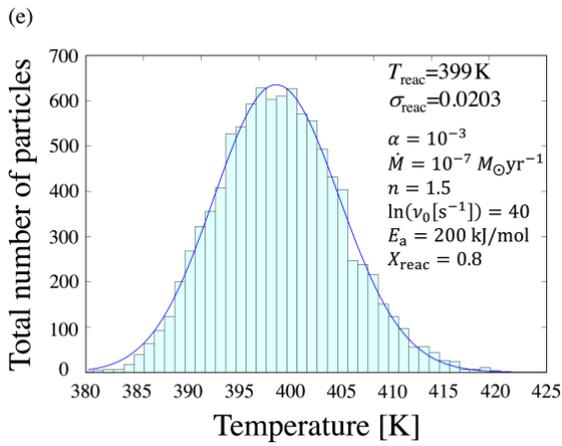
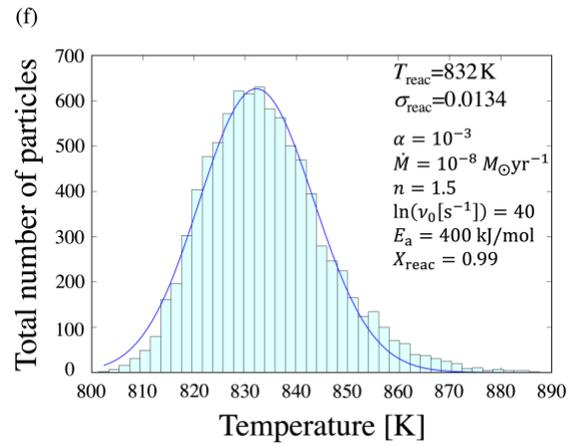

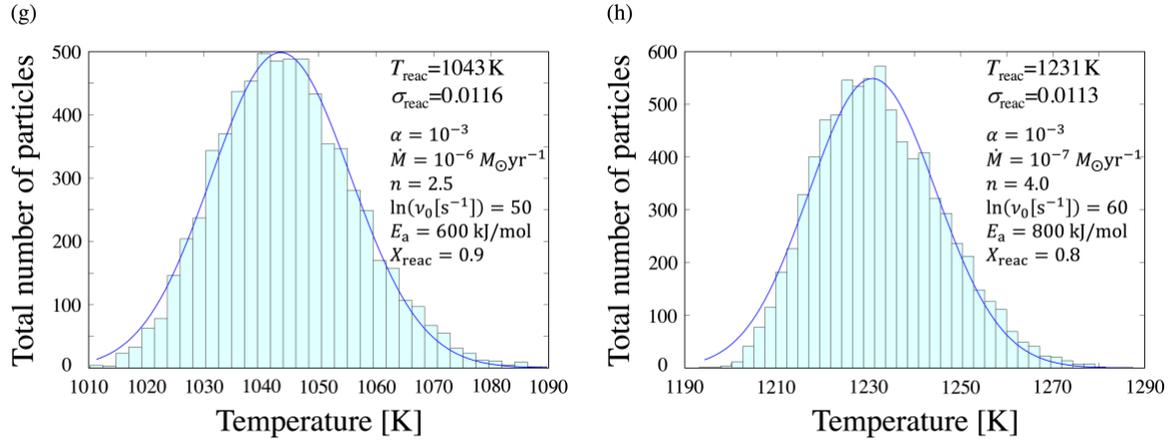

**Figure 2.** Histograms of $T_{\max}^{X_{\text{reac}}}$ (the highest temperature that a dust particle experiences before the reaction degree $X$ reaches $X_{\text{reac}}$) of $10^4$ particles in the simulations with different sets of reaction and disk parameters. The fitted curves with log-normal distributions (Eq.22) are also shown as solid curves.

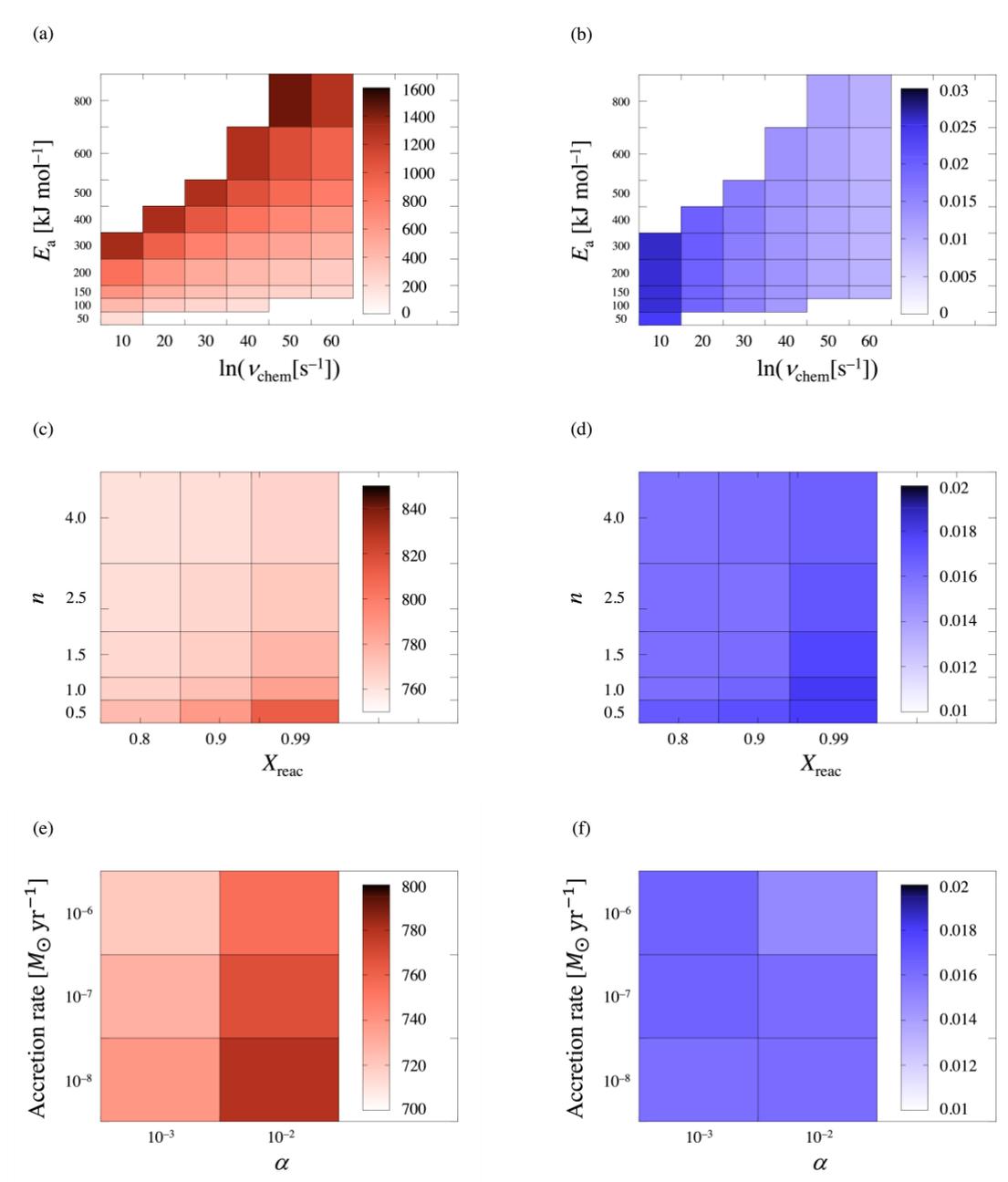

**Figure 3**. $T_{\text{reac}}$ (K) [left panels] and its relative dispersion ($\sigma_{\text{reac}}$) [right panels] for the reactions with different reaction parameter sets ($E_a$, $\ln \nu_{\text{chem}}$, and n) in accretion disks ($\alpha$ and $\dot{M}$). (a) $T_{\text{reac}}$ and (b) $\sigma_{\text{reac}}$ for different $E_a$ and $\ln(\nu_{\text{chem}}[\text{s}^{-1}])$ ($X_{\text{reac}}=0.9$, $n=1.5$, $\alpha=10^{-2}$, and $\dot{M} = 10^{-7}\ M_\odot\ \text{yr}^{-1}$). (c) $T_{\text{reac}}$ and (d) $\sigma_{\text{reac}}$ for different Avrami indexes $n$ and $X_{\text{reac}}$ ($E_a=300$ kJ mol$^{-1}$, $\ln(\nu_{\text{chem}}[\text{s}^{-1}])=30$, $\alpha=10^{-2}$, and $\dot{M} = 10^{-7}\ M_\odot\ \text{yr}^{-1}$). (e) $T_{\text{reac}}$ and (f) $\sigma_{\text{reac}}$ for different disk parameters $\alpha$ and $\dot{M}$ ($X_{\text{reac}}=0.9$, $E_a=300$ kJ mol$^{-1}$, $\ln(\nu_{\text{chem}}[\text{s}^{-1}])=30$, $n=1.5$).

# 4. GENERALIZED FORMULA

## 4.1. Timescales of a chemical reaction and diffusive transport of dust

The reaction of a dust particle in the disk proceeds effectively when the dust particle stays in the region with a certain temperature range, which is high enough for the effective reaction, for a reasonably long duration before they leave the region. Therefore, comparison between a reaction timescale and a local diffusive transport timescale of dust particles is essential to discuss how $T_{\text{reac}}$ (Figs. 2 and 3) is determined. We note that dust movement is driven by both advection and diffusion to the same degree; the advection velocity ($-3\nu_{\text{vis}}/2r$) gives the drift timescale of $\sim 2r^2/3\nu_{\text{vis}}$ that is almost the same as the diffusion timescale of $\sim r^2/\nu_{\text{vis}}$. Therefore, the diffusion timescale can be compared with the reaction timescale as a representative timescale of dust motion.

The reaction timescale is given here as a function of $X$ and $n$ by

$$t_X = \frac{1-X}{|dX/dt|} = \frac{[-\ln(1-X)]^{-(n-1)/n}}{n} \tau = C_X \tau \tag{24}$$

where $\tau$ is the characteristic timescale of a reaction (Eq. 16), and $n$ is the Avrami index (Eq. 15). Equation (24) gives the time required for the completion of reaction from the reaction degree of $X$. $C_X$ is larger than 1 and has positive dependence on $X$ for $n < 1.0$, while it is smaller than 1 with negative $X$ dependence for $n > 1.0$, which will be discussed later when a semi-analytical formula of $T_{\text{reac}}$ is developed (Section 4.2).

The radial diffusion timescale for the radial scale $\Delta r$ is expressed by:

$$t_{\text{diff}} \sim \frac{(\Delta r)^2}{\nu_{vis}} \simeq \alpha^{-1} \left(\frac{\Delta r}{r}\right)^2 \frac{v_K^2}{c_s^2} \Omega_K^{-1}, \tag{25}$$

where the local speed of sound $c_s$ and the Kepler velocity $v_K$ with a solar-mass central star are given by

$$c_s \simeq 1.83 \left(\frac{T_c}{10^3 \text{ K}}\right)^{1/2} \text{ km/s}, \tag{26}$$

$$v_K \simeq 29.8 \left(\frac{r}{1\text{au}}\right)^{-1/2} \text{ km/s}, \tag{27}$$

respectively. Because the midplane temperature is highest at a given $r$ from the Sun and the most effective reaction is expected for the dust moving inward, the midplane temperature $T_c$ is used to evaluate $c_s$.

By substituting Eqs. (26), (27), and (7) into Eq. (25), the diffusion timescale is expressed as:

$$t_{\text{diff}} \simeq 1.03 \times 10^7 \left(\frac{\alpha}{10^{-2}}\right)^{-10/9} \left(\frac{\dot{M}}{10^{-7} M_\odot \text{ yr}^{-1}}\right)^{2/9} \left(\frac{\Delta r/r}{0.01}\right)^2 \left(\frac{T}{10^3 \text{ K}}\right)^{-14/9} \text{ s}. \tag{28}$$

We evaluate $\Delta r/r$ in the following subsections.

### 4.2. Semi-analytical derivation of the effective reaction temperature and its dispersion

We here discuss semi-analytical formulas for the effective reaction temperature and its dispersion to reproduce the numerical results, by comparing the reaction and diffusion timescales of dust. The reaction timescale (Eq. 24) is rewritten as a logarithmic form using Eq. (16):

$$\ln(t_X) = \ln C_X + \ln(\tau \ [s]) = \ln C_X + \frac{E_a}{RT} - \ln(\nu_{chem} \ [s^{-1}]). \tag{29}$$

The diffusion timescale in Eq. (28) is also given as a logarithmic form:

$$\ln(t_{diff}) \simeq 16.15 - \frac{10}{9}\ln\left(\frac{\alpha}{10^{-2}}\right) + \frac{2}{9}\ln\left(\frac{\dot{M}}{10^{-7} M_\odot \ yr^{-1}}\right) + 2\ln\left(\frac{\Delta r/r}{0.01}\right) - \frac{14}{9}\ln\left(\frac{T}{10^3 \ K}\right). \tag{30}$$

At the effective reaction temperature ($T_{reac}$), a chemical reaction proceeds efficiently before the dust diffuses out from the region of $T_{reac}$. From $\ln(t_x) = \ln(t_{diff})$, we obtain the effective reaction temperature of $T_{reac}$ as follows:

$$T_{reac} = \frac{E_a}{R}\left[16.15 - \ln C_X + \ln(\nu_{chem} \ [s^{-1}]) - \frac{10}{9}\ln\left(\frac{\alpha}{10^{-2}}\right)\right.$$
$$\left. + \frac{2}{9}\ln\left(\frac{\dot{M}}{10^{-7} M_\odot \ yr^{-1}}\right) + 2\ln\left(\frac{\Delta r/r}{0.01}\right) - \frac{14}{9}\ln\left(\frac{T}{10^3 \ K}\right)\right]^{-1} K. \tag{31}$$

Equation (31) shows that $T_{reac}$ becomes higher for the larger $X$ when $n < 1$ (Eq. 24), which is consistent with the numerical simulations that show the higher mode temperatures for the larger $X_{reac}$ (Fig. 3). On the other hand, when $n$ is larger than 1, the opposite trend is found for $T_{reac}$ with $C_x$ in Eq. (24). To avoid this inconsistency, we propose to use $C_x$ of $1/n$ for $n > 1$. Because $C_x$ (Eq. 24) does not show a strong dependence on $X$ in the realistic ranges of $n$ ($1 < n < 4$) and $X$ (0.8–0.99), the approximation of $C_x$ as $1/n$ for $n > 1$ is a reasonable assumption. In the discussion below, we use the following $C_x$;

$$C_X = \frac{1}{n}[-\ln(1-X)]^{-(n-1)/n} \quad [\text{for } n < 1], \tag{32}$$

$$C_X = \frac{1}{n} \quad [\text{for } n \geq 1]. \tag{33}$$

Because the variation ranges of disk parameters ($\alpha$, $\dot{M}$, $\Delta r/r$ and $T$) are narrower than those on the reaction parameters ($\nu_{chem}$, and $C_X$) and the $E_a$ dependence is stronger than those of the other parameters, $T_{reac}$ in Eq. (31) has weaker dependence on the disk parameters. Accordingly, a rough estimate of the effective reaction temperature is given by

$$T_0 = \frac{E_a}{R}[16.15 - \ln C_X + \ln(\nu_{chem} \ [s^{-1}])]^{-1} K. \tag{34}$$

Substituting $T_0$ for $T$ in the right-hand side of Eq. (31), $T_{reac}$ is expressed by

$$T_{reac} = \frac{E_a}{R}\left[16.15 - \ln C_X + \ln(\nu_{chem} \ [s^{-1}]) - \frac{14}{9}\ln\left(\frac{T_0}{10^3 \ K}\right) + 2\ln\left(\frac{\Delta r/r}{0.01}\right)\right.$$
$$\left. - \frac{10}{9}\ln\left(\frac{\alpha}{10^{-2}}\right) + \frac{2}{9}\ln\left(\frac{\dot{M}}{10^{-7} M_\odot \ yr^{-1}}\right)\right]^{-1} K. \tag{35}$$

The upper and lower lines of Eq. (35) express the dependences on the reaction parameters and the disk parameters, respectively, because $T_0$ and $\Delta r/r$ are determined only by the reaction parameters (Eqs. 34 and 38). The parameter, $\Delta r/r$, which is not fixed in Eq. (31), can be rewritten using the characteristic reaction timescale (Eq. (16)) and its relative dispersion ($\Delta \tau/\tau$) as

$$\frac{\Delta r}{r} = \left|\frac{\partial \ln r}{\partial \ln T}\frac{\partial \ln T}{\partial \ln \tau}\frac{\Delta \tau}{\tau}\right| = \frac{10}{9}\frac{RT}{E_a}\left|\frac{\Delta \tau}{\tau}\right|, \qquad (36)$$

where the relation,

$$\frac{\Delta T}{T} \simeq \frac{1}{T}\frac{\Delta T}{\Delta r}\Delta r \simeq \left|\frac{d \ln T}{d \ln r}\right|\frac{\Delta r}{r} = \frac{9}{10}\frac{\Delta r}{r}, \qquad (37)$$

is used (Eq. (7)). Applying $T = T_0$ (Eq. 36), $\Delta r/r$ is given as

$$\frac{\Delta r}{r} = \frac{10}{9}\left|\frac{\Delta \tau}{\tau}\right|[16.15 - \ln C_X + \ln(\nu_{\text{chem}}\,[s^{-1}])]^{-1}. \qquad (38)$$

The parameter $\Delta \tau/\tau$ should be empirically determined through comparison with the results of numerical simulations (Section 4.3).

The dispersion of the effective reaction temperature ($\sigma_{\text{reac}}$) is expected to be related to the temperature dispersion within $\Delta r$. Assuming that $\sigma_{\text{reac}}$ is given simply by a linear function of $\Delta T/T$,

$$\sigma_{\text{reac}} = \frac{\Delta T_{\text{reac}}}{T_{\text{reac}}} = C_\sigma \frac{\Delta T}{T} = C_\sigma \left|\frac{\Delta \tau}{\tau}\right|[16.15 - \ln C_X + \ln(\nu_{\text{chem}}\,[s^{-1}])]^{-1}, \qquad (39)$$

Where $C_\sigma$ is a fitting parameter to explain $\sigma_{\text{reac}}$ in the numerical simulations (Fig. 3).

### 4.3. Prediction formulas of the effective reaction temperature and its dispersion

The histograms of $T_{\max}^{X_{\text{reac}}}$ obtained in the numerical simulations are compared with the log-normal distributions using $T_{\text{reac}}$ (Eq. 35) and $\sigma_{\text{reac}}$ (Eq. 39) as a mode temperature (Eq. 23) and its dispersion ($\sigma$ in the log-normal distribution; Eq. (22)), respectively. As described in the previous section, $\Delta\tau/\tau$ (Eq. 36) and $C_\sigma$ (Eq. 39) are the free parameters, while it is most likely that they are of the order of $O(1)$ because the numerical simulations show that the reactions occur in limited temperature ranges depending on the reaction parameters.

Figure 4 compares the histograms of $T_{\max}^{X_{\text{reac}}}$ with predicted distributions of effective reaction temperatures with $\Delta\tau/\tau$ of 0.4 and $C_\sigma$ of 1.8, which were determined by the least absolute value estimation. We found that the predicted distributions of $T_{\text{reac}}$ with these values of $\Delta\tau/\tau$ and $C_\sigma$ well explain the mode temperatures and their dispersion for the reactions in the numerical simulations with wide parameter ranges. The predicted mode temperature ($T_{\text{reac}}$) matches with those in the numerical simulation results within ±5.5% for all the range of reaction and disk parameters (Fig. A1). The relative dispersion of the reaction temperature $\sigma_{\text{reac}}$ also well explain those obtained in the numerical simulations within ±24% (Fig. A2).

We thus conclude that the prediction formulas obtained in this study (Eqs. (35) and (39) with $\Delta\tau/\tau = 0.4$ and $C_\sigma = 1.8$) well reproduce the numerical simulation results for the irreversible reactions occurring in a wide range of disk temperature.

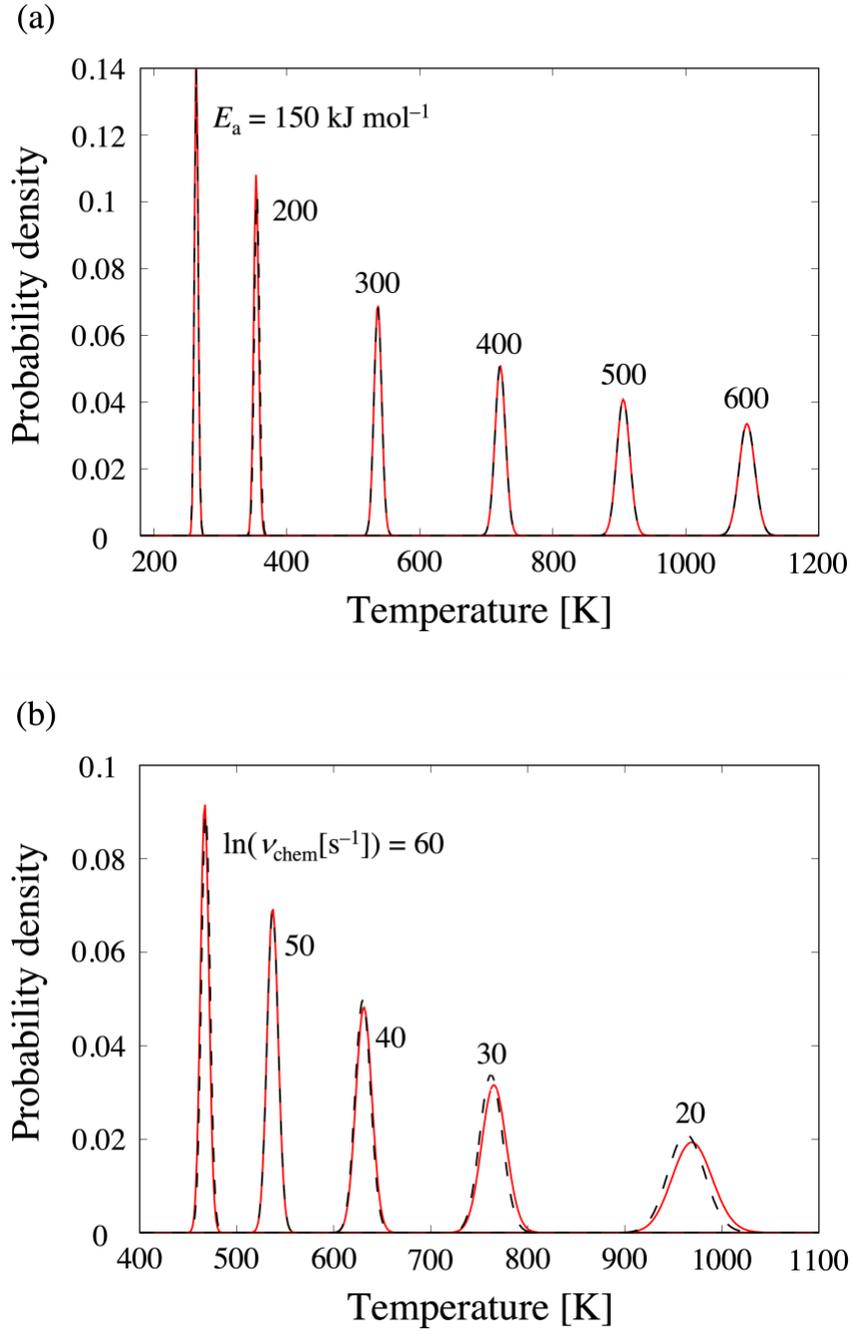

**Figure 4.** The histograms of $T_{\max}^{X_{\mathrm{reac}}}$ ($X_{\mathrm{reac}}$=0.99) obtained in the numerical simulations for various chemical reactions (red solid curves) compared with the predicted distributions (black dashed curves; $\Delta\tau/\tau = 0.4$, $C_\sigma = 1.8$). (a) Reactions with different activation energies ($\ln(\nu_{\mathrm{chem}}[\mathrm{s}^{-1}]) = 50$ and $n = 1.5$). (b) Reactions with different pre-exponential factors ($E_\mathrm{a} = 300$ kJ mol$^{-1}$ and $n = 1.5$). The disk parameters are $\alpha=10^{-2}$ and $\dot{M}=10^{-6}\,M_\odot\,\mathrm{yr}^{-1}$.

# 5. APPLICATION OF PREDICTION FORMULAS FOR CRYSTALLIZATION AND OXYGEN ISOTOPE EXCHANTE OF AMORPHOUS SILICATE DUST AND ITS COSMOCHEMICAL IMPLICATIONS

In this section, we apply the prediction formula to evaluate the temperatures required for the crystallization of amorphous silicate dust and their oxygen isotope exchange with disk $H_2O$ vapor.

It has been pointed out that Earth, Mars, chondrites, differentiated meteorites, asteroids (Itokawa and Ryugu), and a comet Wild 2 are depleted in $^{16}O$ compared to the Sun (e.g., McKeegan et al., 2006, 2011; Yurimoto et al. 2008, 2011; Yokoyama et al., 2022). This difference in oxygen isotopic compositions between the Sun and the Solar System materials has been attributed to the oxygen isotope exchange between silicate dust and $^{16}O$-depleted $H_2O$ vapor, which is likely to have formed through CO self-shielding (Yurimoto & Kuramoto 2004; Lyons & Young, 2005).

Yamamoto et al. (2018, 2020) experimentally determined the oxygen isotope exchange kinetics of amorphous magnesian silicate dust with $H_2O$ vapor, controlled by oxygen isotope diffusion in the amorphous structure. They showed that the oxygen isotope exchange occurs for amorphous silicate dust more effectively than crystalline silicate dust, and discussed that the oxygen isotope exchange should occur at lower temperatures than the crystallization of amorphous silicate for the efficient oxygen isotope exchange. They found that there would be a temperature condition where such effective isotope exchange occurs prior to crystallization, but their discussion did not take the disk dynamics into consideration.

We converted the oxygen isotope exchange rate of sub-micron sized amorphous magnesian silicate dust (80 nm in diameter), having the forsterite ($Mg_2SiO_4$) stoichiometry (hereafter amorphous forsterite), with $H_2O$ vapor in Yamamoto et al. (2018) into the JMA equation ($E_a$ = 161.4 kJ mol$^{-1}$; $\ln(n_{chem}[s^{-1}])$ = 9.80; $n$ = 0.67) to evaluate the effective reaction temperature. We here note that the supply of $H_2O$ may possibly control the exchange reaction at high temperatures (>800 K) and low $H_2O$ pressure (~$10^{-4}$ Pa) for (Yamamoto et al., 2018), but the obtained effective reaction temperatures are <800 K (see below) and thus we did not use the gas supply-controlled rate in this study.

The predicted effective temperature of the oxygen isotope exchange reaction ($X$ = 0.99) and its dispersion are shown in Fig. 5. We also evaluated the effective crystallization temperature ($X$ = 0.99) of amorphous silicate dust with the forsterite stoichiometry using the crystallization kinetics obtained by Yamamoto and Tachibana (2018) ($E_a$ = 414.4 kJ mol$^{-1}$; $\ln(n_{chem}[s^{-1}])$ = 40.2; $n$ = 1.5) (Fig. 5). Yamamoto and Tachibana (2018) concluded that the crystallization of amorphous forsterite in vacuum is a diffusion-controlled reaction within the grain. They found that the reaction parameters change with the partial pressure of $H_2O$ vapor. However, its dependence is not significant enough within the range of disk parameters in this study, and those in vacuum are appropriate to be applied in this work.

We found that crystallization of the amorphous forsterite dust occurs in the temperature range of 800–925 K, while oxygen isotope exchange for 80 nm-sized dust completes at 650–800 K under different sets of disk parameters (Fig. 5). Irrespective of disk parameters, the effective temperatures of two reactions do not overlap each other even considering the reaction temperature dispersions. We also made the 3D Monte Carlo simulation (Section 2) for $10^4$ particles using the reaction parameters above (Yamamoto et al. 2018; Yamamoto & Tachibana 2018) and confirmed that the prediction formulas well represent $T_{reac}$ and $s_{reac}$ obtained in the numerical simulations (Fig. 5).

Yamamoto and Tachibana (2018) found that amorphous forsterite dust crystallizes in $10^6$, $10^4$, $10^2$, and 1 yr at ~700, ~760, ~820, and ~880, respectively, based on the reaction kinetics. Similarly, Yamamoto et al. (2018) showed that 80 nm-sized amorphous forsterite dust can completely exchange oxygen isotope with $H_2O$ vapor for $10^6$, $10^4$, $10^2$, and 1 yr at ~510, ~580, ~675, and ~800 K, respectively. However, the realistic temperature-time ranges of dust in the protoplanetary disk could not be discussed only from the reaction rates. The effective temperatures for the oxygen isotope exchange and crystallization of sub-micron amorphous forsterite in this work are 650–800 and 800–925 K, respectively. These temperature ranges correspond to the characteristic reaction timescales of ~1–100 yr for oxygen isotope exchange and ~0.1–1 yr for crystallization, which can be determined only by comparing the timescales of reactions and dust dynamics in the protoplanetary disk.

The present results (Fig. 5) suggest that sub-micron sized amorphous forsterite dust completes the oxygen isotope exchange with the disk $H_2O$ vapor prior to its crystallization in the accreting disk. Further discussion on the dust grain size, the presence of other gas species (e.g., CO), and the chemical composition of amorphous silicate dust, all of which affect the oxygen isotope exchange reaction of dust (e.g., Yamamoto et al., 2018, 2020), is needed, but this work showed the first quantitative discussion on the oxygen isotope evolution of amorphous silicate dust in the protosolar disk.

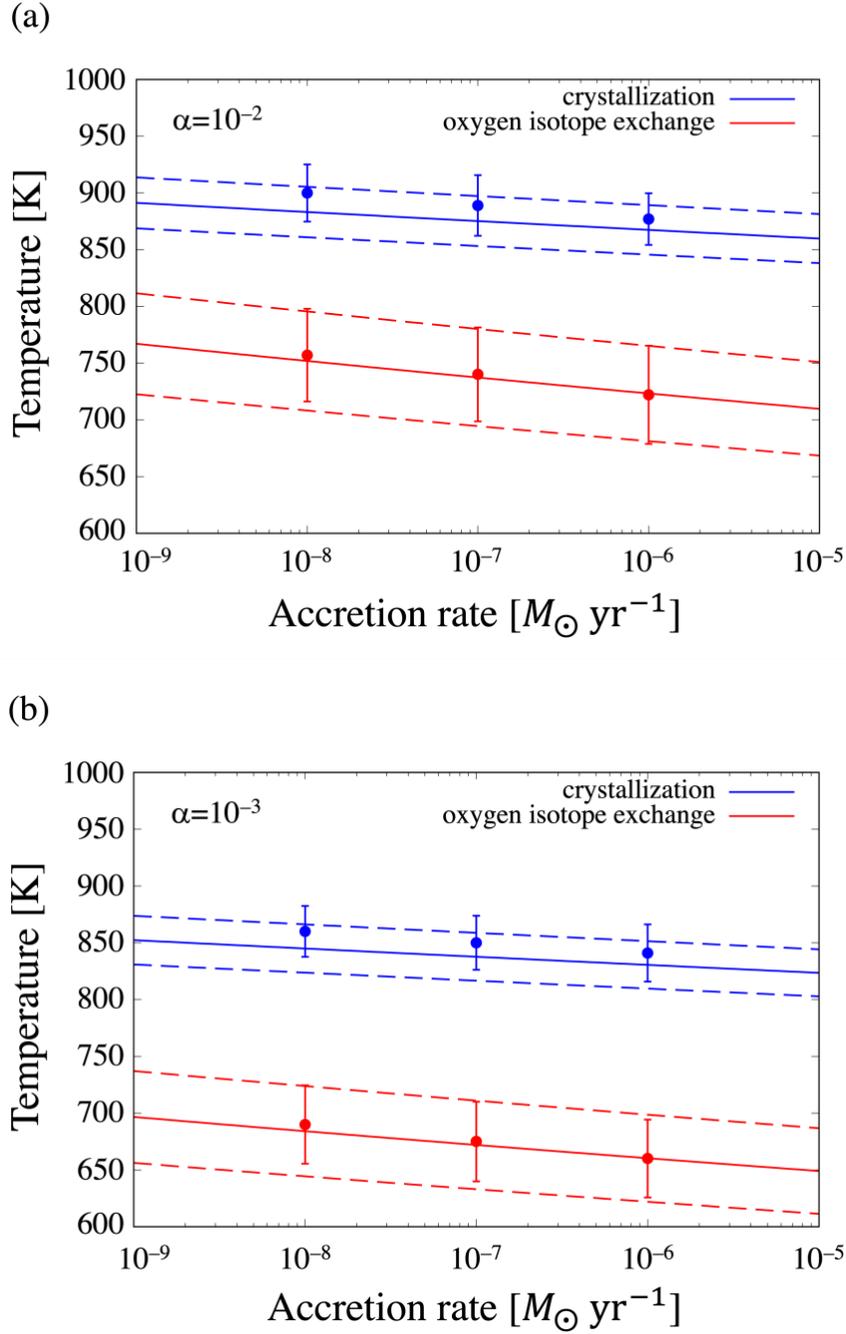

**Figure 5.** The effective reaction temperatures, derived from Eq. (35), for crystallization (blue solid curves) and oxygen isotope exchange (red solid curves) for amorphous forsterite in steady accretion disks with $\alpha=10^{-2}$ and $10^{-3}$. Dispersions of reaction lines (Eq. (39)) are plotted as dashed curves. Solid circles represent $T_{\max}^{X_{\text{reac}}}$ ($X_{\text{reac}}=0.99$) and its dispersion (error bars) obtained by the Monte Carlo simulation (blue: crystallization; red: oxygen isotope exchange).

## 6. CONCLUSIONS

In this study, we performed 3D Monte Carlo simulations of various irreversible chemical reactions of dust particles drifting and diffusing in a steady accretion disk to determine the effective reaction temperatures. We adopt the *r-z* disk temperature determined by viscous heating of which rate is proportional to gas density and inject amorphous submicron-sized silicates at the snow line on the midplane of the disk, following Okamoto and Ida (2022).

The simulations with 5940 parameter sets (6 disk conditions x 330 chemical reactions x 3 reaction degrees) showed that the highest temperatures ($T_{\max}^{X_{\text{reac}}}$) that individual particles experience before the reaction reaches a certain degree show distributions that can be fitted by the log-normal distribution. The mode temperature and the dispersion of the $T_{\max}^{X_{\text{reac}}}$ distribution depend on the reaction and disk parameters, and the mode temperature is defined as the effective reaction temperature.

We developed the semi-analytical prediction formulas of the effective reaction temperature and its dispersion by comparing the reaction timescale with the diffusive transport timescale of dust particles:

$$T_{\text{reac}} = \frac{E_a}{R}\left[16.15 - \ln C_X + \ln(\nu_{\text{chem}}\,[s^{-1}]) + 2\ln\left(\frac{\sigma_{\text{reac}}}{0.0162}\right) - \frac{14}{9}\ln\left(\frac{T_0}{10^3\,\text{K}}\right)\right.$$
$$\left. - \frac{10}{9}\ln\left(\frac{\alpha}{10^{-2}}\right) + \frac{2}{9}\ln\left(\frac{\dot{M}}{10^{-7}M_\odot\,\text{yr}^{-1}}\right)\right]^{-1}\,\text{K}, \qquad (40)$$

where $C_x$ is a parameter related to the reaction (Eqs. 32 and 33) and $T_0$ is determined only by the reaction parameters (Eq. 34). The upper and lower lines of Eq. (40) express the dependences on the reaction parameters and the disk parameters, respectively. The relative dispersion of the reaction line ($\sigma_{\text{reac}}$) is given by

$$\sigma_{\text{reac}} = 0.72 \times [16.15 - \ln C_X + \ln(\nu_{\text{chem}}\,[s^{-1}])]^{-1}. \qquad (44)$$

These prediction formulas of the effective reaction temperature and its dispersion well explain the results of numerical simulations in a wide range of reaction temperatures (~200–1400 K) within ±5.5% and ±24%, respectively.

We applied the prediction formulas for oxygen isotope exchange and crystallization of sub-μm sized amorphous silicate dust with the forsterite composition based on experimentally determined reaction rates. It was found that oxygen isotope exchange of amorphous forsterite dust with $H_2O$ vapor completes before the dust crystallizes in the disk. This would result in effective oxygen isotope evolution of silicate dust in the Solar System and may explain the oxygen isotope difference of planetary materials from the Sun.

These prediction formulas make it possible to discuss various chemical reactions, of which kinetics is obtained in laboratory experiments, in accreting protoplaneraty disks without making numerical simulations, and are a powerful tool to understand distributions of materials that experience different chemical processes in protoplanetary disks.

## 7. ACKNOWLEDGEMENTS

We thank Aki Takigawa for fruitful discussion and two anonymous reviewers for their helpful comments and suggestions. This work is supported by the KAKENHI grant 19H00712, 20H05846, 21H04512, and 21K13986.

## APPENDIX

## DISK VERTICAL TEMPERATURE DISTRIBUTION

With the diffusion approximation, the vertical radiative flux $F$ is given by

$$F = -\frac{4\sigma_B}{3\kappa\rho}\frac{dT^4}{dz},\qquad(A1)$$

where $\sigma_B$ is Stefan-Boltzmann constant. Assuming the optically thick region ($\tau \gtrsim 1$), the flux satisfies

$$\frac{dF}{dz} = q_{\text{vis}} = \frac{9}{4}\rho\nu_{\text{vis}}\Omega_K^2,\qquad(A2)$$

where $q_{\text{vis}}$ is the viscous heating source.

The flux $F$ must be radiated away at $z \sim z_s$. The integration of Eq. (A1) from $z = -z_s$ to $z = z_s$, using approximation of $\int_{-z_s}^{z_s}\rho\, dz \sim \int_{-\infty}^{\infty}\rho\, dz$ because $z_s \sim$ a few $H$, is

$$F(z_s) - F(-z_s) = \int_{-z_s}^{z_s}\frac{9}{4}\rho\nu_{\text{vis}}\Omega_K^2\, dz \simeq \frac{9}{4}\nu_{\text{vis}}\Omega_K^2\int_{-\infty}^{\infty}\rho\, dz = \frac{9}{4}\Sigma_z\nu_{\text{vis}}\Omega_K^2,\qquad(A3)$$

Because $F(z_s) - F(-z_s) = 2F(z_s) = 2s_B T_s^4$ and $S_g\nu_{\text{vis}}W_K^2 = \dot{M}/3\pi$, the photosurface temperature is given by

$$\sigma_B T_{\text{surf}}^4 \simeq \frac{9}{8}\Sigma_z\nu_{\text{vis}}\Omega_K^2 = \frac{9}{8\pi}\dot{M}\Omega_K^2,\qquad(A4)$$

Which is explicitly expressed by Eq. (3):

$$T_{\text{surf}} \simeq 85\left(\frac{\dot{M}}{10^{-8}\,M_\odot/\text{yr}}\right)^{1/4}\left(\frac{M_*}{M_\odot}\right)^{1/4}\left(\frac{r}{1\text{au}}\right)^{-3/4}.\qquad(A5)$$

Integrating Eq. (A1) from $z$ to $z_s$ ($z < z_s$),

$$\int_z^{z_s}\frac{3\kappa F}{4\sigma_B}\rho(z)dz = -\int_z^{z_s}d(T^4) = T(z)^4 - T_{\text{surf}}^4.\qquad(A6)$$

In this paper, the viscous heating is assumed to be proportional to $r(z)$. From Eq. (A6),

$$T(z)^4 - T_{\text{surf}}^4 = \int_z^{z_s}\frac{3\kappa F}{4\sigma_B}\rho(z)dz = -\frac{3}{4\sigma_B}\int_z^{z_s}F(z)d\tau.\qquad(A7)$$

From Eqs. (A2) and (A4) with an assumption that the opacity $k$ does not depend on $z$,

$$\frac{dF}{dz} = -\frac{9\nu_{\text{vis}}}{4\kappa}\Omega_K^2 = -\frac{9\times 3\pi\Sigma_g\nu_{\text{vis}}}{4\times 3\pi\Sigma_g\kappa}\Omega_K^2 = -\frac{3\dot{M}}{4\pi\tau_{\text{disk}}}\Omega_K^2 = -\frac{2\sigma_B}{\tau_{\text{disk}}}T_{\text{surf}}^4.\qquad(A8)$$

Equation (A8) is analytically integrated from 1 to $t_z$,

$$(\tau_z - 1)\frac{2\sigma_B T_{\text{surf}}^4}{\tau_{\text{disk}}} = -\int_1^{\tau_z}\frac{dF}{d\tau}d\tau = -F + \sigma_B T_{\text{surf}}^4.\qquad(A9)$$

Substituting Eq. (A9) into Eq. (A7), we obtain

$$T(\tau_z)^4 - T_{\text{surf}}^4 = \frac{3}{4}\int_1^{\tau_z}T_{\text{surf}}^4\left[1 - \frac{2(\tau_z - 1)}{\tau_{\text{disk}}}\right]d\tau = T_{\text{surf}}^4 \times \frac{3(\tau_z - 1)}{4}\left(1 - \frac{\tau_z - 1}{\tau_{\text{disk}}}\right).\qquad(A10)$$

Therefore, temperature at $z$ is given by Eq. (4):

$$T(z) = \left[1 + \frac{3(\tau_z - 1)}{4}\left(1 - \frac{\tau_z - 1}{\tau_{\text{disk}}}\right)\right]^{1/4}T_{\text{surf}}.\qquad(A11)$$

**DIFFERENCES IN THE PREDICTED REACTION TEMPERATURES AND THEIR DISPERSION FROM THE NUMERICAL SIMULATIONS**

Examples of the deviation of the predicted effective reaction temperatures and their dispersion from those obtained in the numerical reaction is summarized in Figs. A1 and A2.

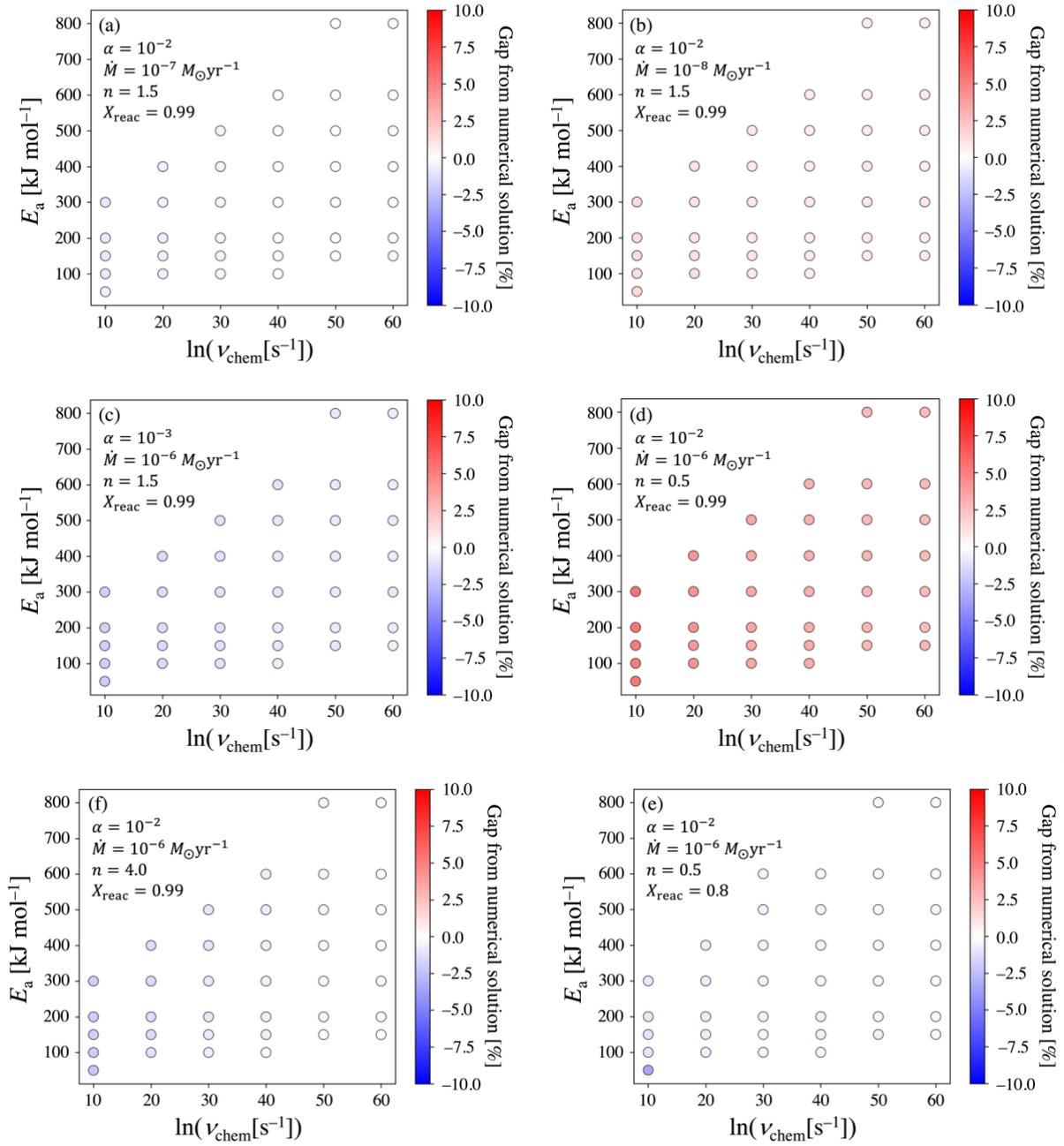

**Figure A1.** Deviation of the predicted effective reaction line temperatures from those obtained in the numerical simulations.

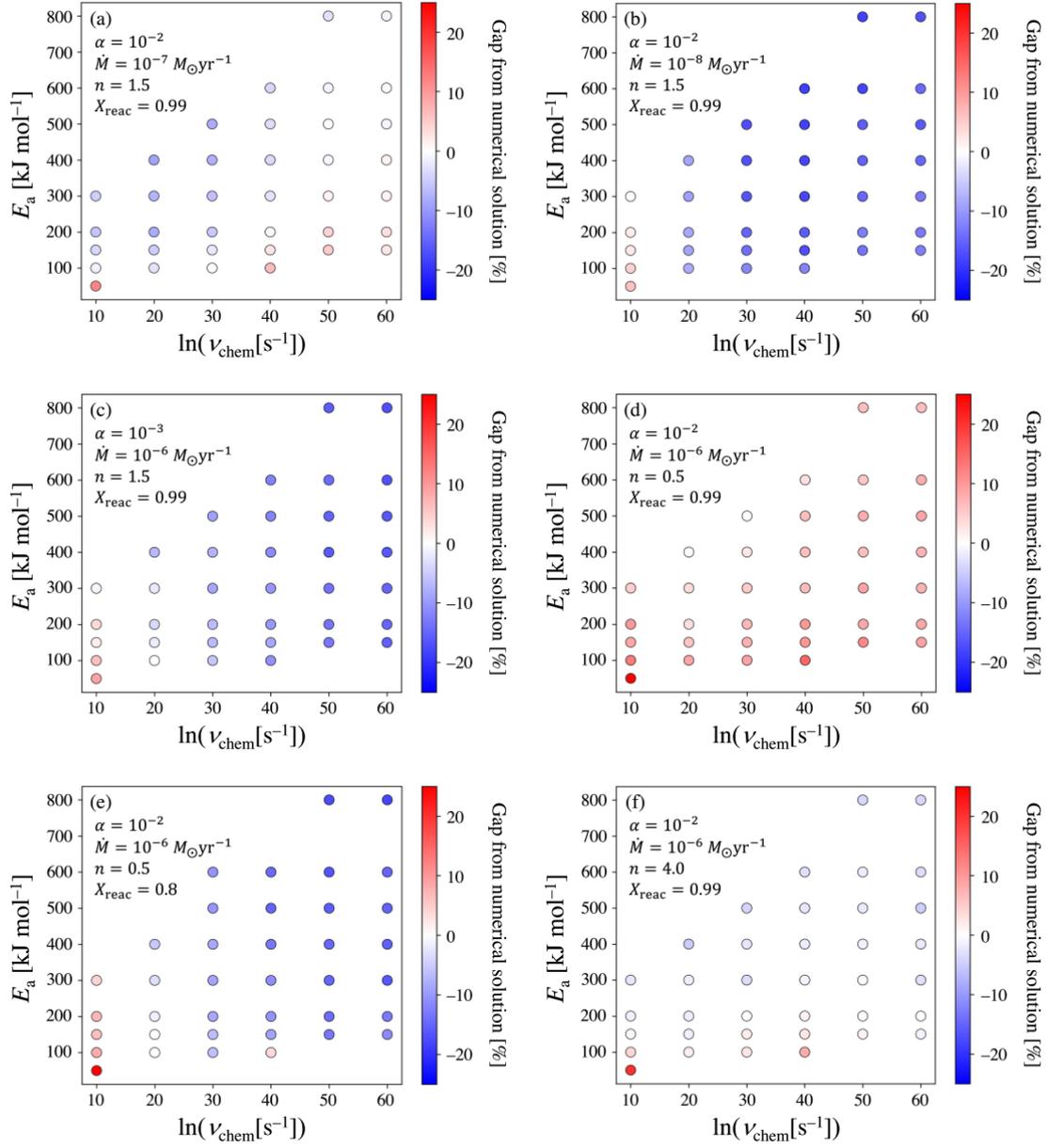

**Figure A2.** Deviation of the relative dispersion of the effective reaction temperatures from those obtained in the numerical simulations.